\documentclass[12pt,preprint,longabstract]{aastex}
\usepackage{array}
\usepackage{longtable}
\usepackage{epsfig}
\usepackage{natbib}
\usepackage{graphicx}
\usepackage{tabularx}
\usepackage{rotating}
\usepackage{multirow}
\usepackage{lscape}
\usepackage{mathrsfs,amssymb}
\usepackage{amsmath}
\usepackage{color}

\shorttitle{Contact Binaries as Viable Near-Infrared Distance Indicators}
\shortauthors{X. Chen et al.}

\begin{document}

\title{Contact Binaries as Viable Distance Indicators: New,
  Competitive $(V)JHK_{\rm s}$ Period--Luminosity Relations}

\author{Xiaodian Chen\altaffilmark{1,2},
        Richard de Grijs\altaffilmark{1,3}, and
        Licai Deng\altaffilmark{2}
        }

\altaffiltext{1}{Kavli Institute for Astronomy \& Astrophysics and
  Department of Astronomy, Peking University Yi He Yuan Lu 5, Hai Dian
  District, Beijing 100871, China; chenxiaodian1989@163.com,
    grijs@pku.edu.cn}
\altaffiltext{2}{Key Laboratory for Optical Astronomy, National
  Astronomical Observatories, Chinese Academy of Sciences, 20A Datun
  Road, Chaoyang District, Beijing 100012, China}
\altaffiltext{3}{International Space Science Institute--Beijing, 1
  Nanertiao, Zhongguancun, Hai Dian District, Beijing 100190, China}

\begin{abstract}
  Based on the largest catalogs currently available, comprising 6090
  contact binaries (CBs) and 2167 open clusters, we determine the
  near-infrared $JHK_{\rm s}$ CB period--luminosity (PL) relations,
  for the first time achieving the low levels of intrinsic scatter
  that make these relations viable as competitive distance
  calibrators. To firmly establish our distance calibration on the
  basis of open cluster CBs, we require that (i) the CB of interest
  must be located inside the core radius of its host cluster; (ii) the
  CB's proper motion must be located within the $2\sigma$ distribution
  of that of its host open cluster; and (iii) the CB's age, $t$, must
  be comparable to that of its host cluster, i.e., $\Delta \log
  (t\mbox{ yr}^{-1}) <0.3$. We thus select a calibration sample of 66
  CBs with either open cluster distances or accurate space-based
  parallaxes. The resulting near-infrared PL relations, for both
  late-type (i.e., W Ursae Majoris-type) and---for the first
  time---early-type CBs, are as accurate as the well-established
  $JHK_{\rm s}$ Cepheid PL relations, (characterized by
    single-band statistical uncertainties of $\sigma < 0.10$ mag). We
  show that CBs can be used as viable distance tracers, yielding
  distances with uncertainties of better than 5\% for 90\% of the 6090
  CBs in our full sample. By combining the full $JHK_{\rm s}$
    photometric data set, CBs can trace distances with an accuracy,
    $\sigma=0.05 \mbox{ (statistical)} \pm0.03 \mbox{ (systematic)}$
    mag. The 102 CBs in the Large Magellanic Cloud are used to
  determine a distance modulus to the galaxy of $(m-M_V)_0^{\rm
    LMC}=18.41\pm0.20$ mag.
\end{abstract}

\section{Introduction}

Contact binaries (CBs), i.e., binary systems where both stellar
components overfill and transfer material through their Roche lobes,
are common among the field stellar population. Their population
density in the solar neighborhood and the Galactic bulge is
approximately 0.2\%, while in the Galactic disk it is, on average,
$\sim$0.1\% \citep{Rucinski06}. CBs are divided into early- and
late-type systems; the latter are also known as W Ursae Majoris (W
UMa) systems. It has been established observationally that the two
binary components have similar temperatures but unequal masses, which
is known as Kuiper's paradox \citep{Kuiper41}. Therefore,
\citet{Lucy68} proposed convective common-envelope evolution as the
key idea of CB theory. The modern scenario is that CBs are formed
through angular-momentum loss \citep[AML;][]{Stepien06, Yildiz13}.

Although CBs are some seven magnitudes fainter than Cepheid variables,
within the same distance range their number is three orders of
magnitude larger. Unlike Cepheids, however, which trace young ($\la
20$ Myr-old) features, CBs map 0.5--10 Gyr-old stellar
populations. Although RR Lyrae stars also trace structures older than
1--2 Gyr, very few RR Lyrae have been found in either open clusters
(OCs) or the solar neighborhood.

Since \citet{Eggen67}'s seminal work, these considerations have led to
a number of attempts at using CB period--luminosity (PL)--color (PLC)
relations as potential distance indicators. \citet{Rucinski94}
obtained a (widely used) PLC relation based on 18 W UMa-type
CBs. \citet{Rucinski97} improved their PLC relation based on 40 W
UMa-type CBs using {\sl Hipparcos} parallaxes with an accuracy in the
corresponding distance moduli of $\epsilon_{M}<0.5$
mag. \citet{Rucinski06} subsequently derived a CB luminosity function
from the All Sky Automated Survey (ASAS) and explored the viability of
a $V$-band PL relation. However, his PL relation exhibited only a weak
correlation and was affected by large uncertainties.

Despite a large volume of new data, studies advocating W UMa-type CB
PL relations as viable distance indicators have made little progress
during the past decade. In addition to calibration inhomogeneities
stemming from difficulties in dealing with extinction and the use of
multiple studies based on a few objects at a time, the current impasse
is predominantly caused by difficulties in distinguishing foreground
and background CBs from genuine OC members. Since CBs represent old
stellar populations, they need a comparably long formation timescale,
so for a typical old OC with 500 stellar members, one or more of its
CBs could be misidentified as a cluster member.\footnote{Support for
  this estimate is provided by \citet{Chen16}, who assessed stellar
  contamination in the low-density OC NGC 188. Of 910 stars detected
  in a $20\times20$ arcmin$^2$ region centered on the cluster, 532 are
  cluster members, while one foreground CB was found among the 378
  contaminants.} In denser environments, the potential contamination
by field CBs is even more significant.

To address these difficulties conclusively, here we introduce a
joined-up OC--CB analysis based on carefully considered proper-motion
and age-selection criteria \citep{Chen15}. We have collected the
largest CB sample currently available. We thus obtain the first truly
viable near-infrared (NIR) CB PL relations, which are comparably
accurate as the well-established $JHK_{\rm s}$ Cepheid PL relations.

In Section 2, we summarize the theory underlying the CB PL
relations. The method used to select OC CBs and obtain OC distances is
discussed in Section 3. The resulting $JHK_{\rm s}$ and $V$-band PL
relations, as well as the corresponding period--color relations, are
explored in Section 4. We discuss CBs as distance tracers and
determine the distance to the Large Magellanic Cloud (LMC) in Section
5, followed by a summary of our main conclusions in Section 6.

\section{Theoretical framework}

From the equations governing contact binaries,
\begin{equation}
\label{equation10}
   \begin{aligned}
   &G(m_1+m_2)=(2\pi/P)^2A^3, \\
   &A \equiv (R/R_{\odot})/r, \\
   &L/L_{\odot}=(T_{\rm eff}/T_{{\rm eff},\odot})^4(R/R_{\odot})^2, \mbox{ and} \\
   &M_{\rm bol}=M_V+{\rm BC},
   \end{aligned}
\end{equation}
a relationship between the distance modulus and other physical
parameters (expressed in solar units) follows trivially,
\begin{equation}
   \label{equation1}
   \begin{aligned}
   (V_{\rm max}-M_{V})=&-39.189+V_{\rm max}+{\rm BC}+10 \log T_1+5 \log r \\
   &           + \frac{5}{3}\log m_1+\frac{10}{3} \log P+\frac{5}{3} \log (1+q),
   \end{aligned}
\end{equation}
where $V_{\rm max}$, $M_{V}$, and $M_{\rm bol}$ are the maximum and
absolute $V$-band magnitudes and the bolometric magnitude of the
binary system, respectively, `BC' is the bolometric correction, $T_1,
T_2, m_1$, and $m_2$ are the temperatures and masses of both
components, $q \equiv m_2/m_1$ is the system's mass ratio, $P$ its
orbital period, $A$ and $R$ are the semi-major axis and the
  stellar radius (assuming blackbody radiation, which is an incorrect
  assumption for realistic stellar atmosphered but which has very
  little influence on our infal results), $r_1$ and $r_2$ are the
  radii of both components in units of the semi-major axis, and the
  relative radius $r$, in the same units, is defined as
\begin{equation}
   \label{equation2}
   r=\left(r_1^2+r_2^2\left(\frac{T_2}{T_1}\right)^4\right)^{\frac{1}{2}}.
\end{equation}
Since $m_1$ and $q$ depend on the effective temperature, $T_{\rm
  eff}$, \citet{Rucinski94} suggested that the terms, $5\log
r+\frac{5}{3} \log(1+q)+\frac{5}{3} \log(m_1)$, may be
omitted. Indeed, for 100 well-studied CBs with known parameters
\citep{Yildiz13}, an obvious linear relationship is apparent between
$5\log r+\frac{5}{3} \log(1+q)+\frac{5}{3} \log(m_1)$ and $\log T_{\rm
  eff}$. Therefore, a simplified PLC relation is indeed well
established, taking the form $L=a\log P+ b\log(T_{\rm eff})+$ constant
(where $a$ and $b$ are constants). By reference to the Cepheid PLC and
PL relations, W UMa-type CBs are thus expected to follow a PL relation
if they also exhibit linear period--color or luminosity--color
relations. Similarly as for Cepheids, the resulting CB PL relations
are more easily obtained at NIR wavelengths.

\section{Period--Luminosity Relations}

To establish the NIR PL relations, independent distance determinations
are needed. First, we selected 20 of the 21 W UMa-type CBs with {\sl
  Hipparcos} parallaxes characterized by distance-modulus
uncertainties of $\epsilon_{M}<0.25$ mag \citep{Rucinski06}. The
remaining system, TY Men, was excluded, since its light curve shows
obvious unequal maxima, a signature of the O'Connell effect. Second,
we collected a total of 6090 CBs with $10 \le V \le 14$ mag, including
1131 and 5374 CBs from the General Catalog of Variable Stars (GCVS)
and the ASAS Catalog of Variable Stars, respectively. In addition, we
used the latest version of the DAML02 OC catalog \citep[][version 2016
  January]{Dias02}, which contains 2167 OCs.

Next, we considered individual publications focusing on CBs in a
number of OCs, including NGC 188 \citep{Zhang04}, NGC 1245
\citep{Pepper06}, NGC 2099 \citep{Kang07}, NGC 2126 \citep{Liu09}, NGC
2158 \citep{Mochejska04}, NGC 2204 \citep{Rozyczka07}, NGC 2243
\citep{Kaluzny06}, NGC 2301 \citep{Kim01}, NGC 2539 \citep{Choo03},
NGC 2682 \citep{Yakut09}, NGC 5381 \citep{Pietrzynski97}, NGC 6253
\citep{De Marchi10}, NGC 6259 \citep{Ciechanowska06}, NGC 6705
\citep{Koo07}, NGC 6791 \citep{De Marchi07}, NGC 6819
\citep{Street02}, NGC 6866 \citep{Molenda-Zakowicz09}, NGC 6939
\citep{Maciejewski08}, NGC 7044 \citep{Kopacki08}, NGC 7142
\citep{Sandquist11}, NGC 7789 \citep{Mochejska99}, Berkeley 39
\citep{Mazur99}, Collinder 261 \citep{Mazur95}, and Melotte 66
\citep{Zloczewski07}. These OC CBs contribute mostly to the faint end
($V>14$ mag) of the CB luminosity function.

To obtain a sample of high-probability OC CBs and reduce
foreground/background contamination, we applied three selection
criteria \citep{Chen15}: (i) the CB of interest must be located inside
the core radius of its host OC
\citep{Dias02,Kharchenko13,Kharchenko16}; (ii) the CB's proper motion
must be located within the $2\sigma$ distribution of that of its host
OC; and (iii) the CB's age, $t$, must be comparable to that of its
host OC, where one typically adopts $\Delta \log (t\mbox{
    yr}^{-1}) <0.3$ \citep[e.g.,][]{Anderson13}. We adopted proper
motions from the Fourth US Naval Observatory CCD Astrograph Catalog
\citep[UCAC4;][]{Zacharias13}, complemented with measurements taken
from the PPMXL Catalog \citep{Roeser10}. Proper-motion selection is
highly efficient in removing foreground CBs.

The evolution of CBs can roughly be divided into the detached,
semi-detached, and contact phases. The mechanism that drives this
  evolution is thought to include Kozai cycles, accompanied by tidal
  friction (KCTF) and AML \citep{Stepien06,
    Yildiz13}. \citet{Tokovinin06} found that 96\% of field binaries
  with orbital periods $P < 3$ days include a third component. This
  third component tends to have a high inclination and can shorten the
  binary's orbital period through KCTF. The associated timescale is
  very short, approximately 50 Myr \citep{Eggleton06}. This mechanism
  becomes ineffective when the orbital period becomes shorter than
  approximately two days \citep{Eggleton06}. \citet{Fabrycky07} also
  found that the KCTF process dominates binaries with periods $2.0 < P
  < 3.0$ days. When the orbital period drops to less than 2 days, AML
  become effective. AML through magnetic winds is also called
  `magnetic braking;' its timescale is longer than 1 Gyr. Therefore,
  only the AML timescale is considered here. The timescale governing
the detached phase is almost the same as the time spent on the main
sequence, which is determined by the initial mass of the primary
component. The modern scenario that CBs are formed through AML can
explain many observations, including:
\begin{enumerate}
\item CBs are preferentially found in intermediate-age or old OCs --
  e.g., NGC 188: 5 Gyr \citep{Chen16}; Berkeley 39: 6 Gyr
  \citep{Mazur99}; Collinder 261: 6 Gyr \citep{Mazur95} -- and they
  are very rare in young, $\sim$1 Gyr-old OCs. Pre-CB counterparts
  (semi-detached and detached binaries) are found in young OCs
  \citep{Rucinski98, Rucinski00}.

\item the density of CBs is 0.2\% in the Galactic bulge and the solar
  neighborhood, but it decreases to 0.1\% in the Galactic disk
  \citep{Rucinski06}. In relatively young environments, the
    fraction of CBs is low, while in old(er) environments this
    proportion increases. This means that the CB formation timescale
    is very long, comparable to the AML and nuclear-evolution
    timescales.

\item the CBs' short-period limit. \citet{Rucinski07} used the
  ASAS catalog to conclude that the short-period limit of CBs is 0.22
  days. There may be a number of reasons for the appearance of such a
  limit: (i) CBs with effective temperatures below the full-convection
  point are dynamically unstable \citep{Rucinski92}; (ii) the
  formation timescale for short-period CBs is longer than the long AML
  timescale \citep{Stepien06}; and/or (iii) mass transfer in low-mass
  CBs is unstable and these CBs can only exist for very short periods
  \citep{Jiang12}. However, following \citet{Rucinski07}'s
  publication, 367 ultra-short-period binaries, with periods of less
  than 0.22 days, have been found in the Catalina survey
  \citep{Drake14}. This survey focuses on sources located away from
  the Galactic disk. Upon application of our PL relations (see below)
  to 202 ultra-short-period CBs with Sloan Digital Sky Survey colors,
  we find that 200 are in located the thick disk or halo, with only
  two residing in the Galactic thin disk. In addition, in the Galactic
  bulge the ultra-short-period limit to the CB distribution is around
  0.19 days \citep{Soszynski15}. Many shorter-period CBs are found in
  very old environments. This suggests that the long AML timescale may
  be the main reason for the short-period limit of CBs, since only
  this mechanism is directly related to the evolutionary timescale. We
  also speculate that the formation timescale of ultra-short-period
  (0.19 days) CBs may be close to the Hubble time.

\item the ultra-short period of detached binaries is around 0.09 days,
  which is significantly shorter than that for CBs (0.19 days)
  \citep{Soszynski15}. Detached binaries are less evolved,
    whereas at least one component of CBs has evolved quite
    significantly (all hydrogen has been converted into helium in the
    stellar core). Ultra-short-period binaries are composed of two
    low-mass components; low-mass stars are characterized by longer
    nuclear-evolution timescales. This suggests that CB formation is
    limited by the long timescales of both nuclear evolution and AML.

\item the period--age relation of CBs and semi-detached binaries
  \citep{Bukowiecki12}. Although period--age relations are not
  reliable for single CBs, a mean age--period relation (pertaining to
  a given cluster) may apply. If so, short periods are equivalent to
  lower luminosities (lower masses) and older ages.

\end{enumerate}

OCs represent, to first approximation, single stellar
populations. This means that all stellar members have the same
age. The pre-CB evolutionary timescale must be shorter than their host
open cluster's age ($t_{\rm cb} < t_{\rm cl}$) in order for these
binaries to become CBs. This suggests that high-mass eclipsing
binaries can evolve into CBs on short timescales. We obtain the
  limiting mass, $M_{1,{\rm limit}}$, of the initial primary
  components from the limiting evolutionary timescale, $t_{\rm cb} =
  t_{\rm cl}$. \citet{Yildiz13} have derived an expression for $t_{\rm
    cb}$ based on stellar evolution models, i.e.,
\begin{equation}
   \label{equation8}
{{\text{t}}_{\text{cb}}}\text{ = }\frac{\text{10}}{{{\text{(}M\text{/}{{\text{M}}_{\odot }}\text{)}}^{\text{4}\text{.05}}}}\times \left( \text{5}\text{.60}\cdot \text{1}{{\text{0}}^{\text{-}3}}{{\left( \frac{M}{{{\text{M}}_{\odot }}}\text{+}3.993 \right)}^{3.16}}+0.042 \right)
\end{equation}

\begin{figure}
\centering
\includegraphics[width=120mm]{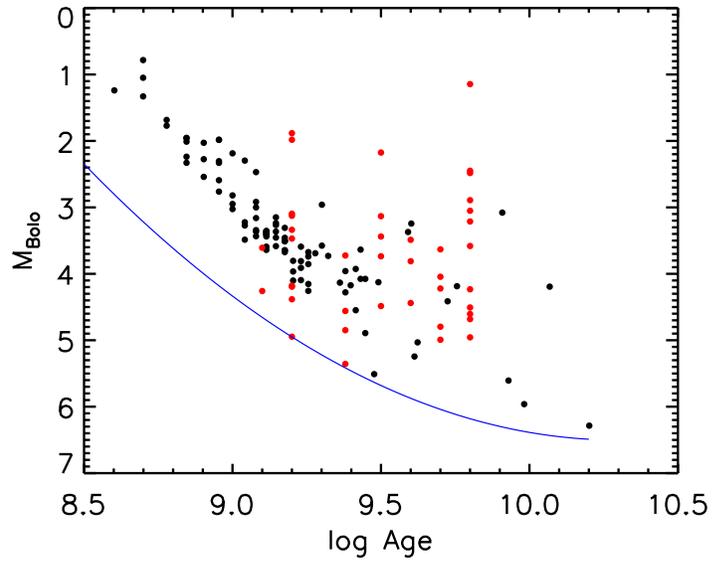}
\caption{Age selection. Black dots: 100 well-studied CBs
  \citep{Yildiz13}; solid line: lower limit to the ages of these CBs,
  characterized by $\Delta \log (t\mbox{ yr}^{-1}) =0.3$; red dots:
  final sample of 42 OC CBs.}
  \label{f1.fig}
\end{figure}

The pre-CB counterparts satisfying $m_{1,i}>m_{1, {\rm limit}}$ can
evolve into OC CBs. Based on this $m_{1,{\rm limit}}$ and different
$m_{2,i}$ values, we can estimate the lower limit to the current
masses and luminosities of our sample CBs. CBs with luminosities below
this limiting luminosity are likely background stars. This
(conservative) age-selection criterion (see Fig. \ref{f1.fig}) is
based on 100 well-studied CBs from \citet{Yildiz13}. Age selection is
very effective in excluding background CBs. We checked whether the
resulting PL relations would exhibit reduced scatter if we adopted
tighter age constraints, but we did not find any difference. Among our
sample OC CBs, two are characterized by $0.2 \le \Delta \log (t\mbox{
  yr}^{-1}) <0.3$ and three have $0.1 \le \Delta \log (t\mbox{
  yr}^{-1}) <0.2$. However, these five CBs are affected by large
uncertainties and low weights in establishing the PL relations.

After application of all three selection criteria, 42 high-probability
OC CBs remain. Next, we determined CB distances and reddening values
based on their host clusters' properties. We collected the most recent
results for every OC (see Table \ref{t1}) and checked our results
using NIR data from the Two Micron All Sky Survey (2MASS), where
available.\footnote{This analysis was done for NGC 188
  \citep{Hills15}, NGC 2158 \citep{Carraro02}, NGC 2184
  \citep{Kharchenko16}, NGC 2243 \citep{Anthony05}, NGC 2682
  \citep{Geller15}, NGC 6705 \citep{Santos05}, NGC 6791
  \citep{Carney05}, NGC 6819 \citep{Balona13}, NGC 6939
  \citep{Andreuzzi04}, NGC 7044 \citep{Sagar98}, NGC 7142
  \citep{Straizys14}, NGC 7789 \citep{Wu07}, Berkeley 39
  \citep{Bragaglia12}, and Collinder 261 \citep{Gozzoli96}.} We used
the main-sequence fitting technique based on the Padova isochrones
\citep{Girardi00}, applied to the proper-motion-selected cluster
members to estimate the distances and reddening values \citep{Chen15}.

\section{Results}

Our final CB sample consists of 42 OC CBs, four nearby moving-group
CBs with well-determined distances, and 20 W UMa-type CBs with
high-accuracy {\sl Hipparcos} parallaxes: see Table \ref{t1}. Since
the scatter in the PL relations decreases from $V$ to $K_{\rm s}$
\citep[e.g.,][]{Madore91}, we focus on the NIR PL relations. To
establish the latter, we need access to the maximum $JHK_{\rm s}$
luminosities. Since the light-curve shapes do not change significantly
as a function of wavelength compared with, e.g., Cepheid variables
(because the variations are caused by eclipses and the resulting BCs
for the two binary components are negligible, given that they have
similar temperatures), the maximum magnitude in a given band can be
obtained by converting from single-epoch magnitudes to maximum
luminosities using well-established light curves in the $V$ band
(taken from the literature). To reduce the effects of changes in the
orbital period in our photometric conversions, for each CB we adopted
the closest primary minimum epoch, $T_0$, to the observed 2MASS
epoch. The maximum uncertainties of this conversion include 10\%
of total amplitude and 10\% of phase uncertainty. The uncertainties in
the absolute magnitudes (see Table \ref{t1}) are a combination of the
distance-modulus error, the photometric error, the extinction error,
and these two conversion errors.

\begin{center}
\begin{tiny}
\begin{longtable}{lcccccclccc}
\caption{Calibration sample of 66 CBs. $M_J(\varphi)$, $M_H(\varphi)$,
  and $M_{K_s}(\varphi)$ are the absolute magnitudes at phase
  $\varphi$, which can be converted to maximum magnitudes using
  $m_{\rm max}=m_{\varphi}-m_{\rm adj}$. DM (distance modulus),
  $E(J-H)$, and $\log( t )$ are the best parameters for each
  OC.}\label{t1}  \\
  \hline
  Contact Binary  & Period & $M_J(\varphi)$ & $M_H(\varphi)$ & $M_{K_s}(\varphi)$ & Phase ($\varphi$) & $m_{\rm adj}$ & OC & DM  &$E(J-H)$ & $\log( t )$      \\
          & (day)    & (mag)        & (mag)          & (mag)             &                & (mag)   &          & (mag)    &(mag)    &  [year]       \\
  \hline
   V782 Cep            & 0.3583 & 3.145(161)  & 2.784(159)  & 2.661(164)  & 0.83 & 0.05(01)  & NGC 188       & 11.29(10)  & 0.03(01)  & 9.7\\
  ep cep              & 0.2897 & 3.882(200)  & 3.443(225)  & 3.205(241)  & 0.94 & 0.35(04)  & NGC 188       & 11.29(10)  & 0.03(01)  & 9.7\\
  eq cep              & 0.3070 & 4.013(210)  & 3.489(241)  & 3.372(255)  & 0.10 & 0.40(04)  & NGC 188       & 11.29(10)  & 0.03(01)  & 9.7\\
  V370 Cep            & 0.3304 & 3.164(177)  & 2.879(178)  & 2.793(182)  & 0.62 & 0.08(01)  & NGC 188       & 11.29(10)  & 0.03(01)  & 9.7\\
  es cep              & 0.3425 & 2.770(157)  & 2.521(512)  & 2.292(508)  & 0.27 & 0.01(01)  & NGC 188       & 11.29(10)  & 0.03(01)  & 9.7\\
  AH CNC              & 0.3605 & 2.813(148)  & 2.571(146)  & 2.525(144)  & 0.89 & 0.20(02)  & NGC 2682      & 9.55 (10)  & 0.01(00)  & 9.5\\
  EV CNC              & 0.4414 & 2.225(133)  & 1.935(126)  & 1.891(125)  & 0.16 & 0.04(01)  & NGC 2682      & 9.55 (10)  & 0.01(00)  & 9.5\\
  HS CNC              & 0.3597 & 2.844(139)  & 2.535(136)  & 2.508(133)  & 0.97 & 0.10(01)  & NGC 2682      & 9.55 (10)  & 0.01(00)  & 9.5\\
  NU CMa              & 0.2853 & 3.244(274)  & 2.795(283)  & 2.382(519)  & 0.92 & 0.14(01)  & NGC 2243      & 13.15(10)  & 0.02(01)  & 9.2\\
  NW CMa              & 0.3565 & 2.681(196)  & 2.460(233)  & 2.118(284)  & 0.61 & 0.14(01)  & NGC 2243      & 13.15(10)  & 0.02(01)  & 9.2\\
  V521 Lyr            & 0.3257 & 3.052(233)  & 2.938(292)  & 2.448(514)  & 0.33 & 0.04(01)  & NGC 6791      & 13.09(10)  & 0.04(01)  & 9.6\\
  V513 Lyr            & 0.3917 & 2.783(207)  & 2.445(234)  & 2.377(515)  & 0.20 & 0.05(01)  & NGC 6791      & 13.09(10)  & 0.04(01)  & 9.6\\
  J19203636+3739567   & 0.3664 & 3.269(252)  & 2.752(292)  & 2.467(325)  & 0.44 & 0.20(02)  & NGC 6791      & 13.09(10)  & 0.04(01)  & 9.6\\
  V2388 Cyg           & 0.3660 & 2.805(230)  & 2.524(225)  & 2.429(243)  & 0.33 & 0.03(01)  & NGC 6819      & 11.87(14)  & 0.05(01)  & 9.4\\
  V2396 Cyg           & 0.2932 & 3.542(245)  & 3.131(248)  & 2.910(294)  & 0.95 & 0.20(02)  & NGC 6819      & 11.87(14)  & 0.05(01)  & 9.4\\
  V2393 Cyg           & 0.3032 & 3.592(257)  & 3.042(576)  & 2.597(568)  & 0.05 & 0.15(02)  & NGC 6819      & 11.87(14)  & 0.05(01)  & 9.4\\
  V2394 Cyg           & 0.2562 & 4.052(256)  & 3.504(260)  & 3.439(563)  & 0.66 & 0.10(01)  & NGC 6819      & 11.87(14)  & 0.05(01)  & 9.4\\
  NGC 6939 MGN V20    & 0.2951 & 3.747(281)  & 3.439(278)  & 3.232(289)  & 0.99 & 0.50(05)  & NGC 6939      & 11.41(10)  & 0.11(03)  & 9.1\\
  NGC 6939 MGN V6     & 0.3599 & 2.536(203)  & 2.191(173)  & 2.052(175)  & 0.18 & 0.02(01)  & NGC 6939      & 11.41(10)  & 0.11(03)  & 9.1\\
  KDK2008 v5          & 0.6150 & 1.576(503)  & 1.411(450)  & 1.170(432)  &      & 0.23(12)  & NGC 7044      & 12.47(20)  & 0.20(06)  & 9.2\\
  KDK2008 v6          & 0.6547 & 1.228(451)  & 1.011(393)  & 0.994(374)  &      & 0.13(07)  & NGC 7044      & 12.47(20)  & 0.20(06)  & 9.2\\
  KDK2008 v3          & 0.4606 & 2.261(505)  & 1.924(462)  & 1.866(472)  &      & 0.21(11)  & NGC 7044      & 12.47(20)  & 0.20(06)  & 9.2\\
  NGC 7789 KP V7      & 0.3375 & 3.146(309)  & 2.771(292)  & 2.661(299)  & 0.49 & 0.12(01)  & NGC 7789      & 11.42(20)  & 0.07(02)  & 9.2\\
  V875 Cas            & 0.3063 & 3.430(337)  & 3.064(336)  & 2.959(341)  & 0.15 & 0.37(04)  & NGC 7789      & 11.42(20)  & 0.07(02)  & 9.2\\
  J18510018$-$0614494 & 0.8696 & 0.569(331)  & 0.474(318)  & 0.493(558)  & 0.33 & 0.05(03)  & NGC 6705      & 11.33(20)  & 0.14(02)  & 8.4\\
  J21451515+6549242   & 0.5808 & 1.564(220)  & 1.384(193)  & 1.319(171)  & 0.94 & 0.07(01)  & NGC 7142      & 11.98(10)  & 0.12(03)  & 9.5\\
  J21442843+6546365   & 0.3302 & 3.370(215)  & 2.730(192)  & 2.736(175)  & 0.00 & 0.36(01)  & NGC 7142      & 11.98(10)  & 0.12(03)  & 9.5\\
  J06071751+2404455   & 0.3555 & 2.608(274)  & 2.459(264)  & 2.385(281)  &      & 0.10(04)  & NGC 2158      & 13.06(10)  & 0.13(03)  & 9.2\\
  J06074059+2405035   & 0.3635 & 2.704(420)  & 2.399(387)  & 2.100(435)  &      & 0.24(05)  & NGC 2158      & 13.06(20)  & 0.13(03)  & 9.2\\
  V705 Mon            & 0.3810 & 2.560(465)  & 2.225(390)  & 2.551(444)  & 0.87 & 0.05(09)  & Berkeley 39   & 12.92(20)  & 0.03(03)  & 9.8\\
  V711 Mon            & 0.3063 & 3.338(298)  & 2.653(316)  & 2.645(426)  & 0.08 & 0.15(01)  & Berkeley 39   & 12.92(20)  & 0.03(01)  & 9.8\\
  V712 Mon            & 0.2844 & 3.364(352)  & 2.978(350)  & 3.329(621)  & 0.30 & 0.03(01)  & Berkeley 39   & 12.92(20)  & 0.03(01)  & 9.8\\
  V704 Mon            & 0.2780 & 3.949(340)  & 2.933(387)  & 2.263(616)  & 0.12 & 0.30(01)  & Berkeley 39   & 12.92(20)  & 0.03(01)  & 9.8\\
  V706 Mon            & 0.4865 & 2.004(433)  & 1.680(384)  & 1.922(628)  & 0.56 & 0.16(02)  & Berkeley 39   & 12.92(20)  & 0.03(01)  & 9.8\\
  V701 Mon            & 0.5419 & 1.848(287)  & 1.661(314)  & 1.623(364)  & 0.64 & 0.10(02)  & Berkeley 39   & 12.92(20)  & 0.03(01)  & 9.8\\
  V938 Mon            & 0.6820 & 1.023(278)  & 0.966(286)  & 0.802(324)  & 0.63 & 0.16(01)  & Berkeley 39   & 12.92(20)  & 0.03(01)  & 9.8\\
  J23571065+5633268   & 0.2790 & 3.644(279)  & 3.142(281)  & 3.284(272)  &      & 0.05(01)  & NGC 7789      & 11.42(20)  & 0.07(01)  & 9.2\\
  IR Mus              & 0.3158 & 3.768(357)  & 3.440(400)  & 2.355(598)  & 0.64 & 0.22(02)  & Collinder 261 & 12.09(15)  & 0.08(03)  & 9.8\\
  HM Mus              & 0.3431 & 3.523(346)  & 3.216(348)  & 2.904(606)  & 0.13 & 0.30(03)  & Collinder 261 & 12.09(15)  & 0.08(03)  & 9.8\\
  HU Mus              & 0.5190 & 1.902(259)  & 1.722(241)  & 1.598(238)  & 0.47 & 0.10(01)  & Collinder 261 & 12.09(15)  & 0.08(03)  & 9.8\\
  HZ Mus              & 0.4295 & 2.255(275)  & 2.034(242)  & 1.980(256)  & 0.74 & 0.00(01)  & Collinder 261 & 12.09(15)  & 0.08(03)  & 9.8\\
  IO Mus              & 0.4009 & 2.480(300)  & 2.172(318)  & 2.113(299)  & 0.03 & 0.28(03)  & Collinder 261 & 12.09(15)  & 0.08(03)  & 9.8\\
  ASAS061214$-$0347.4 & 0.3637 & 2.871(204)  & 2.581(197)  & 2.579(192)  & 0.20 & 0.06(01)  & NGC 2184      & 8.96 (15)  & 0.03(01)  & 8.7\\
  ASAS081347$-$4034.2 & 0.5831 & 1.719(266)  & 1.692(255)  & 1.653(243)  & 0.07 & 0.14(01)  & Ruprecht 56   & 8.15 (20)  & 0.04(01)  & 8.8\\
  TX Cnc              & 0.3829 & 2.867(147)  & 2.591(151)  & 2.528(141)  & 0.10 & 0.20(02)  & praesepe      & 6.16 (10)  & 0.01(00)  & 8.9\\
  QX And              & 0.4122 & 2.175(130)  & 2.017(128)  & 1.961(125)  & 0.16 & 0.03(01)  & NGC 752       & 8.19 (10)  & 0.01(00)  & 9.2\\
  VW Cep              & 0.2783 & 3.759(123)  & 3.222(102)  & 3.109(090)  & 0.01 & 0.42(04)  &               &       &      &    \\
  OU Ser              & 0.2968 & 3.294(124)  & 3.030(144)  & 2.918(120)  & 0.14 & 0.05(01)  &               &       &      &    \\
  SX Crv              & 0.3166 & 3.122(270)  & 2.855(284)  & 2.780(256)  & 0.37 & 0.03(01)  &               &       &      &    \\
  YY Eri              & 0.3215 & 3.112(244)  & 2.775(251)  & 2.669(238)  & 0.14 & 0.12(01)  &               &       &      &    \\
  W UMa               & 0.3336 & 2.935(204)  & 2.585(209)  & 2.546(192)  & 0.32 & 0.04(01)  &               &       &      &    \\
  GM Dra              & 0.3387 & 2.636(205)  & 2.384(198)  & 2.341(192)  & 0.71 & 0.02(01)  &               &       &      &    \\
  V757 Cen            & 0.3432 & 3.353(171)  & 3.049(199)  & 2.940(179)  & 0.90 & 0.22(02)  &               &       &      &    \\
  V781 Tau            & 0.3449 & 2.749(235)  & 2.463(231)  & 2.405(229)  & 0.67 & 0.08(01)  &               &       &      &    \\
  GR Vir              & 0.3470 & 3.506(169)  & 3.264(169)  & 3.220(174)  & 0.92 & 0.40(04)  &               &       &      &    \\
  AE Phe              & 0.3624 & 3.054(147)  & 2.765(144)  & 2.658(147)  & 0.11 & 0.19(02)  &               &       &      &    \\
  YY CrB              & 0.3766 & 2.813(208)  & 2.579(230)  & 2.482(214)  & 0.42 & 0.25(03)  &               &       &      &    \\
  V759 Cen            & 0.3940 & 2.726(127)  & 2.484(143)  & 2.421(127)  & 0.03 & 0.20(02)  &               &       &      &    \\
  EX Leo              & 0.4086 & 2.223(236)  & 2.006(248)  & 1.938(243)  & 0.37 & 0.06(01)  &               &       &      &    \\
  V566 Oph            & 0.4096 & 2.171(142)  & 2.007(152)  & 1.961(146)  & 0.22 & 0.01(01)  &               &       &      &    \\
  AW UMa              & 0.4387 & 2.087(152)  & 1.948(148)  & 1.916(154)  & 0.73 & 0.02(01)  &               &       &      &    \\
  CN Hyi              & 0.4561 & 1.922(102)  & 1.716(115)  & 1.638(103)  & 0.75 & 0.00(01)  &               &       &      &    \\
  RR Cen              & 0.6057 & 1.799(186)  & 1.686(206)  & 1.608(183)  & 0.46 & 0.35(04)  &               &       &      &    \\
  IS CMa              & 0.6170 & 1.571(137)  & 1.420(140)  & 1.349(136)  & 0.15 & 0.15(02)  &               &       &      &    \\
  V535 Ara            & 0.6293 & 1.649(277)  & 1.500(294)  & 1.479(276)  & 0.51 & 0.49(05)  &               &       &      &    \\
  S Ant               & 0.6484 & 1.454(114)  & 1.327(126)  & 1.232(102)  & 0.35 & 0.11(01)  &               &       &      &    \\
  \hline
  \end{longtable}
\end{tiny}
\end{center}

\begin{figure}
\centering
\includegraphics[width=120mm]{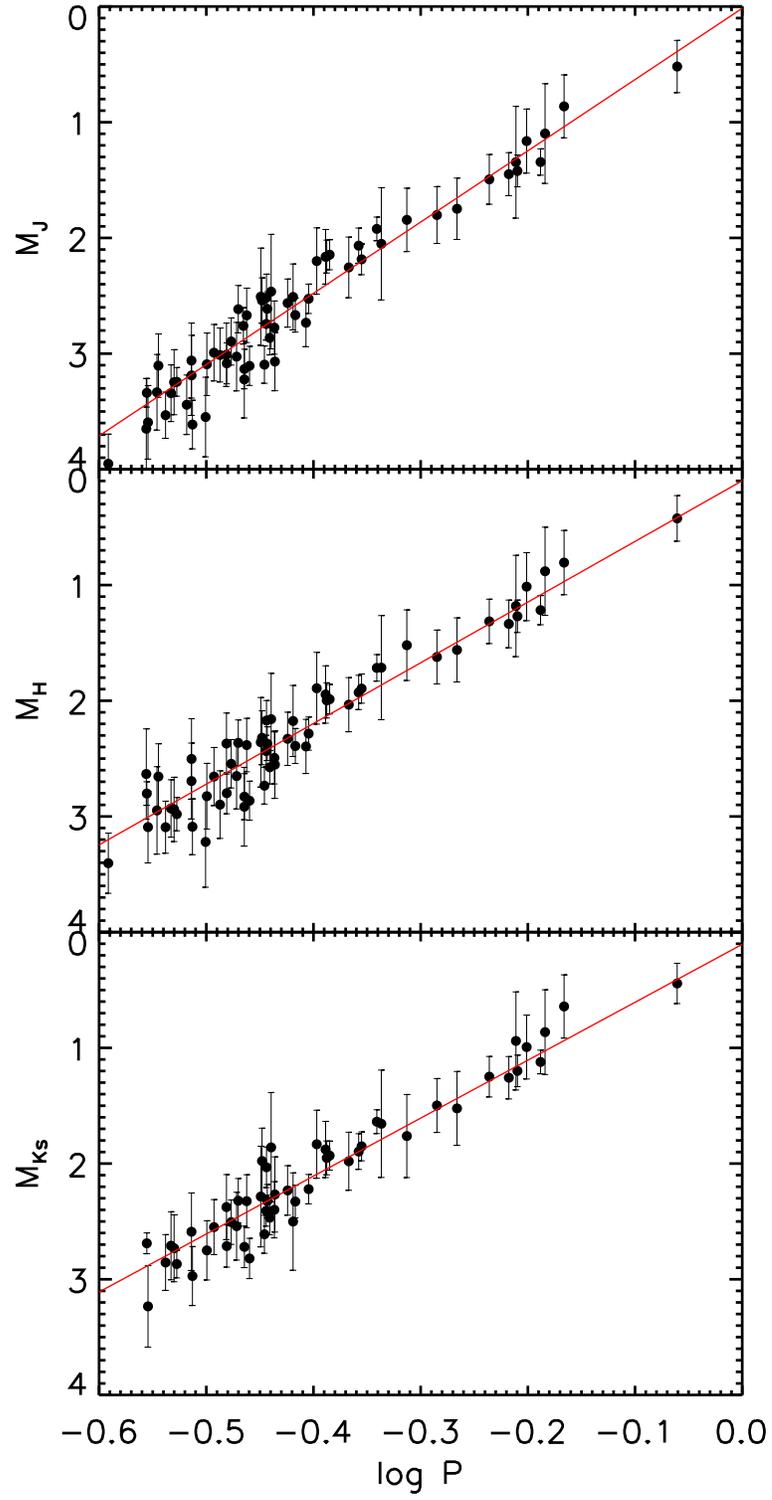}
\caption{$JHK_{\rm s}$ PL relations determined based on our 66
  calibration CBs.}\label{f4.fig}
\end{figure}

Based on the 66 CBs in our calibration sample we obtain
\begin{equation}\label{equation3}
  \begin{aligned}
   M_{J_{\rm max}}^{\rm late} &= (-6.15 \pm 0.13) \log P+(-0.03 \pm 0.05), \sigma_J=0.09, \log P < -0.25; \\
   M_{J_{\rm max}}^{\rm early} &= (-5.04 \pm 0.13) \log P+(0.29 \pm 0.05), \sigma_J=0.09, \log P > -0.25; \\
   M_{H_{\rm max}} &= (-5.22 \pm 0.12) \log P+(0.12 \pm 0.05), \sigma_H=0.08; \\
   M_{K_{\rm s,max}} &= (-4.98 \pm 0.12) \log P+(0.13 \pm 0.04), \sigma_K=0.08.
   \end{aligned}
\end{equation}
These PL relations---shown in Fig. \ref{f4.fig}---are as accurate as
the $JHK_{\rm s}$ PL relations for Cepheids based on OC distances
\citep{Chen15}, which means that CBs can indeed be statistically
competitive distance tracers to old stellar populations.

RR Lyrae also trace $\ga 1$--2 Gyr-old stellar populations. However,
their density is much lower than that of CBs. For instance, in the
magnitude-limited ASAS survey, the number of RR Lyrae with $V < 14$
mag is 1450. This is only a quarter of the number of late-type CBs in
our sample (5374) and similar to the numbers of both Cepheids (1182)
and early-type CBs (1582). Some long-period Cepheids may not have been
detected. However, RR Lyrae periods are in the range 0.2--1.0 days,
i.e., similar to those of CBs.

The magnitudes of RR Lyrae range from $K_{\rm s} = -0.5$ mag to
$K_{\rm s} = +0.5$ mag; CB magnitudes range from $K_{\rm s} = -0.6$
mag to $K_{\rm s} = 4$ mag (late-type CBs: $1 \la K_{\rm s} \la 4$
mag). Early-type CBs ($-0.6 \le K_{\rm s} \le 1.0$ mag) are as bright
as RR Lyrae and their number is roughly the same. CBs have as
advantage that they trace intermediate-age environments (4--6 Gyr),
while RR Lyrae represent the first choice to determine distances to
old environments (older than 10 Gyr), such as old globular clusters.

The $J$-band PL relation is only reliable for late (W UMa)-type CBs; a
small correction is needed for early-type CBs (for $\log P > -0.25$;
Rucinski 2006). The adjustment in the $H$ band is less than the
uncertainty; no correction is needed in the $K_{\rm s}$ band. In
optical bands, the differences between both types of CBs are more
obvious on account of the period--color relation. \citet{Rucinski06}
treated CBs with periods $\log P> -0.25$ as early-type CBs and
excluded them from his derivation of the late-type CB PL
relation. Compared with late-type CBs, thus far the early-type CB
PL(C) relation has been poorly studied, since the number of early-type
CBs with accurate distance estimates has been limited. The theory
underlying the early-type CB PL relation is not well-established
compared with that pertaining to late-type CBs. Based on their
periods, we infer that the initial primary components of early-type
CBs, $m_{1,i}$, have higher masses compared to their counterparts in
late-type CBs. For early-type CBs, when both components fill their
Roche lobes, $m_1$ is still larger than $m_2$, while late-type CBs
undergo a mass reversal that leads to $m_1<m_2$ when both components
fill their Roche lobes. Early-type CB have longer periods and
  need shorter evolutionary timescales.

\begin{figure}
\includegraphics[width=85mm]{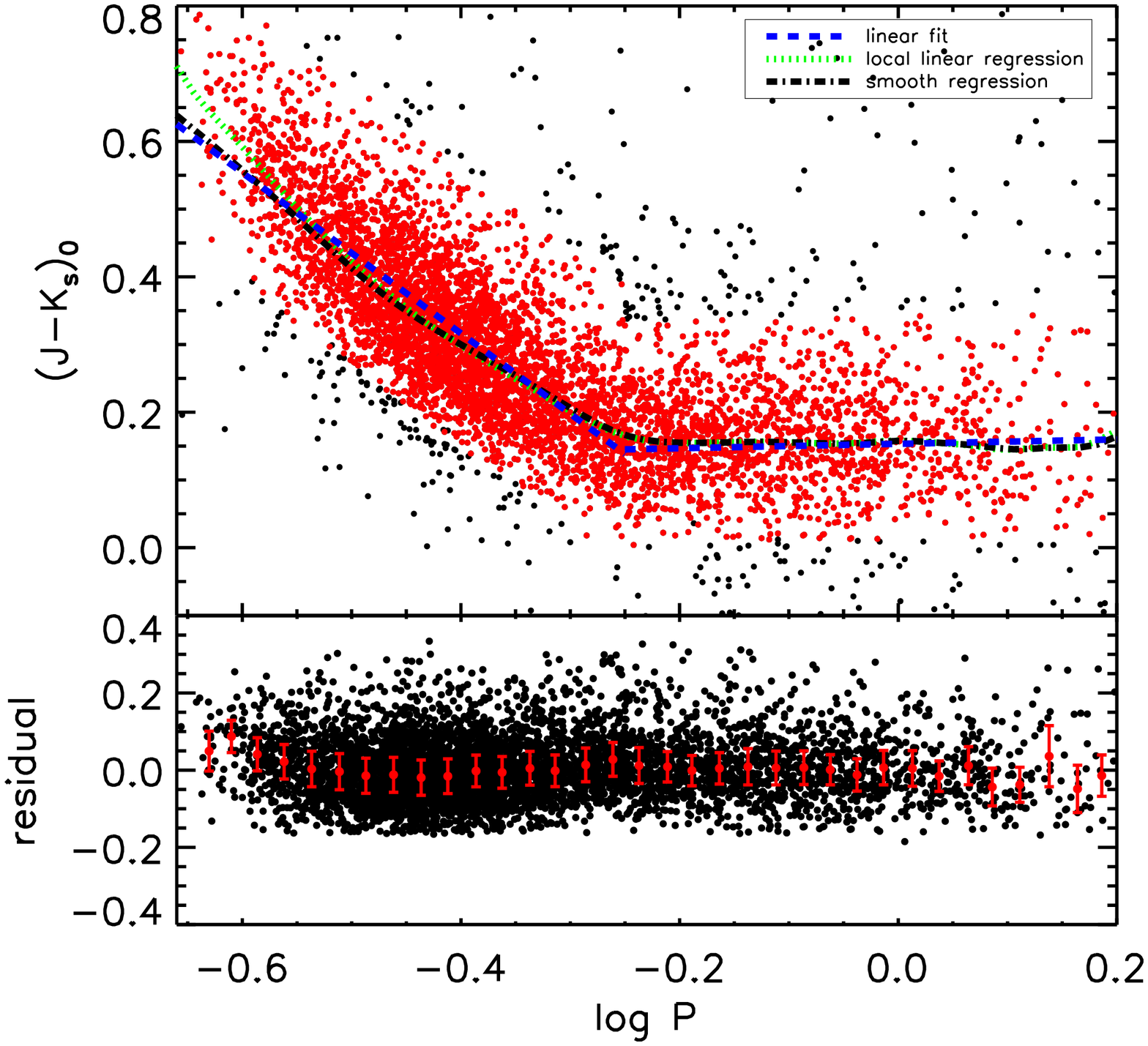}
\includegraphics[width=85mm]{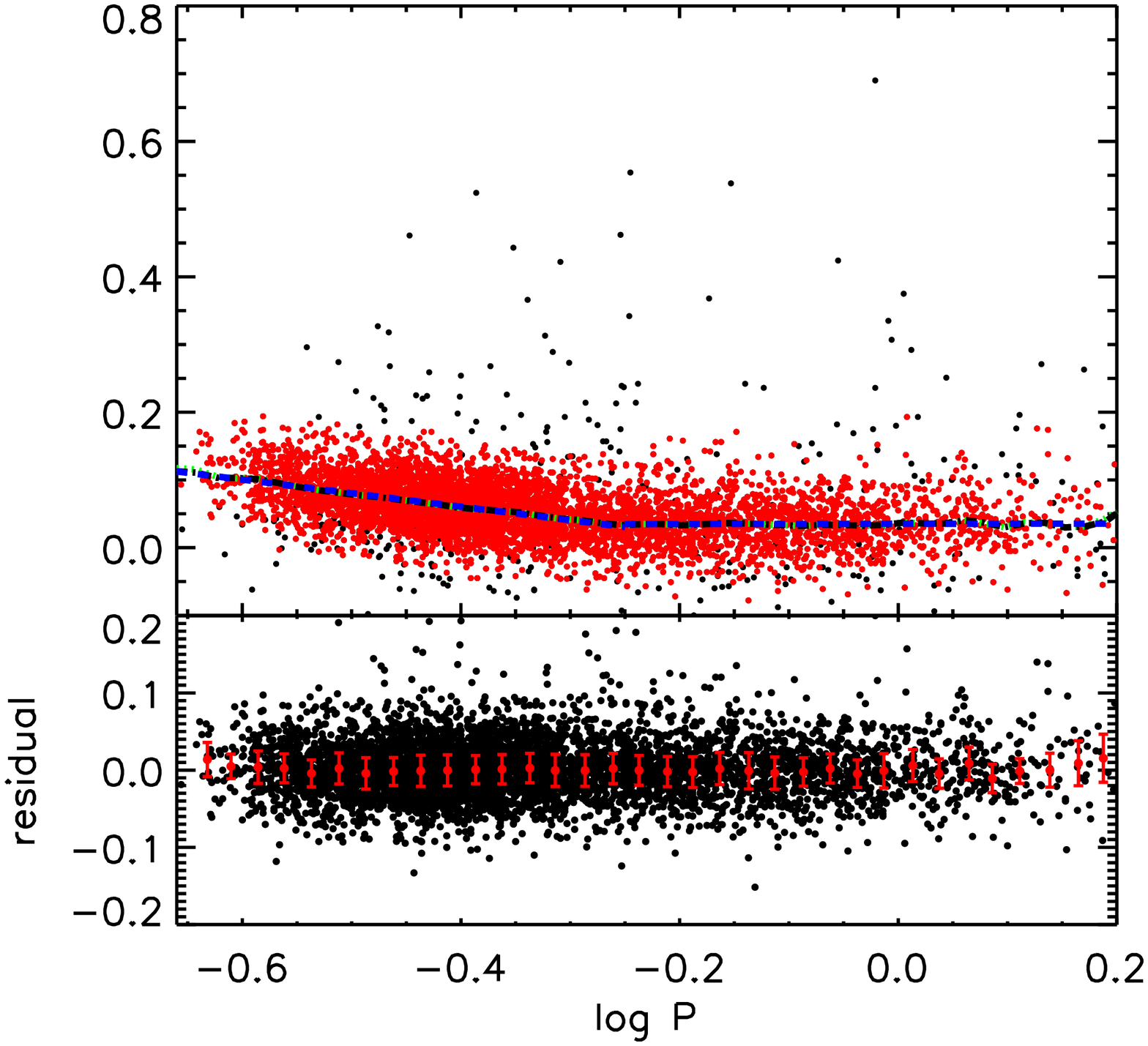}
\includegraphics[width=85mm]{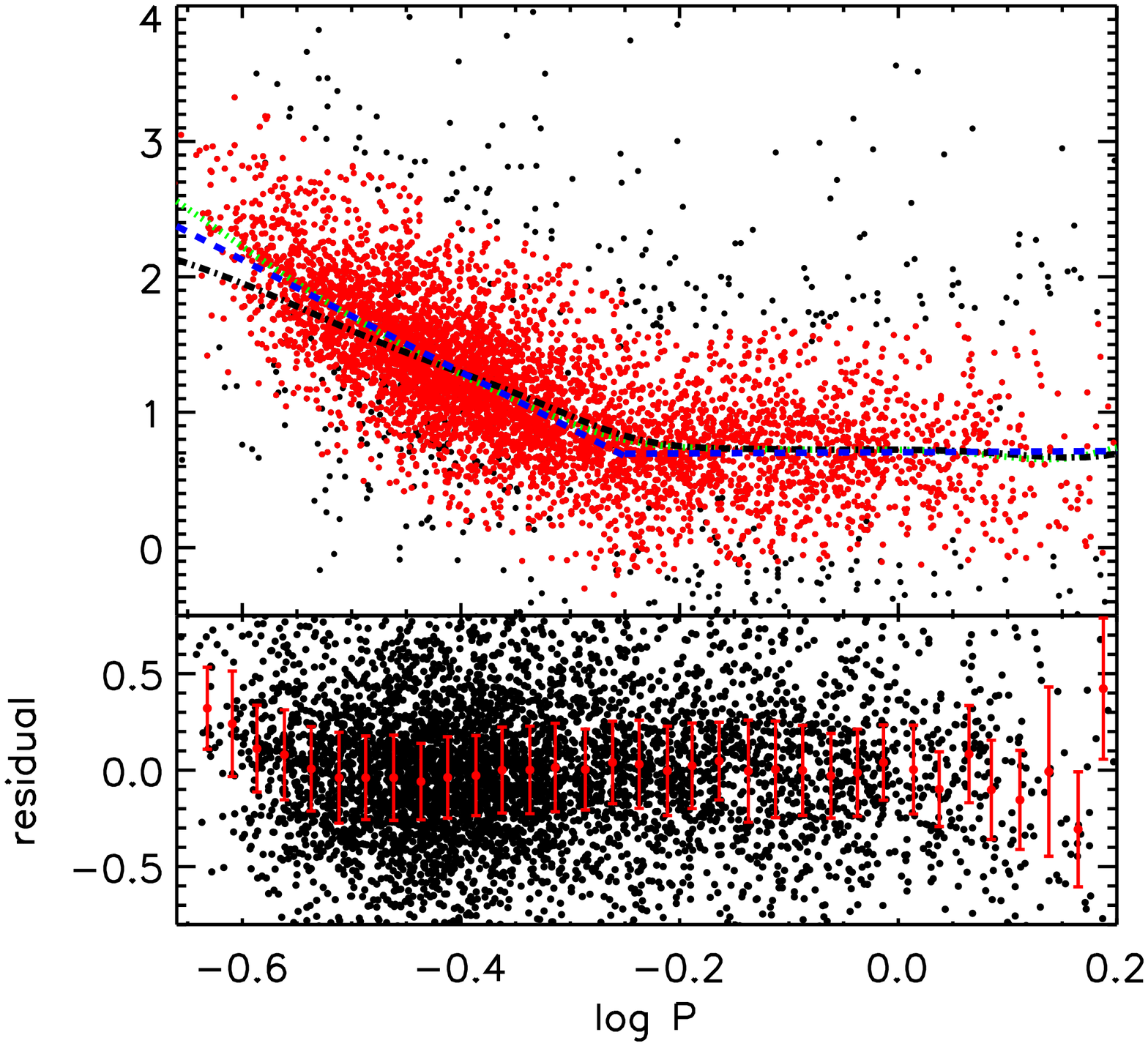}
\caption{Period--color relations for our full sample of 6090 CBs.
  Early- and late-type CBs each follow different period--color
  relations. The best fits are shown as solid lines. Red dots: CBs
  located within the 3$\sigma$ distributions. Black dots: CBs with
  poor-quality photometry, affected by complicated differential
  extinction, or objects that are not genuine CBs. The residuals are
  also shown for each relation. Green dotted lines are based on
    local linear regression, while black dash-dotted lines are based
    on smooth regression. Although the $J-K_{\rm s}$ and $V-K_{\rm
    s}$ diagrams may exhibit small trends at the short- and
  long-period extremes, these only cover a few tens of CBs and may not
  be real. Indeed, they also disappear in the less reddening-sensitive
  $H-K_{\rm s}$ diagram.}\label{f5.fig}
\end{figure}

\begin{figure}
\centering
\includegraphics[width=120mm]{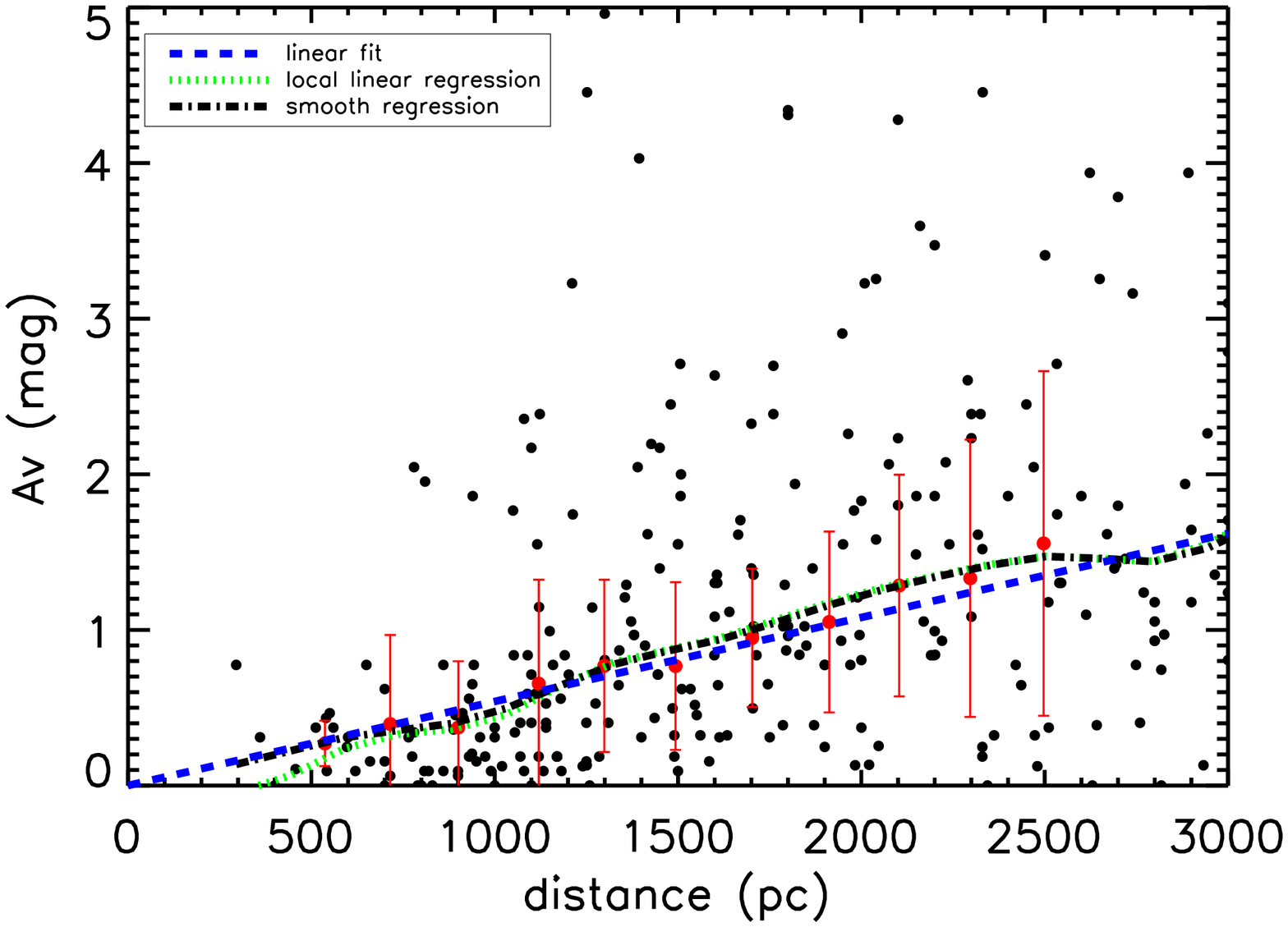}
\caption{$V$-band extinction of 445 old OCs obtained from the DAML02
  catalog \citep{Dias02}. The blue circles are the average extinction
  values in each bin at distances from 400 pc to 2600 pc. The red
  dashed line is the statistical average extinction, $A_V=0.54\pm
  0.23$ mag kpc$^{-1}$, based on all black points. The green
    dotted and black dash-dotted lines are based on local linear
    regression and smooth regression, respectively.}\label{f6.fig}
\end{figure}

The period--color relation can help us quantify the accuracy when
converting PLC to PL relations. Figure \ref{f5.fig} shows the NIR
period--color relations for our full sample of 6090 CBs, adjusted
  for extinction corresponding to $A_V=(0.54\pm 0.23)$ mag kpc$^{-1}$
  (see Fig. \ref{f6.fig}). We derived the latter trend from a
  statistical analysis of all 445 OCs with ages greater than $10^9$ yr
  in the DAML02 catalog. The extinction per kiloparsec distance was
  derived for every OC; average values could then be obtained
  directly. A 1$\sigma$ cut was applied to exclude OCs with unusually
  deviating extinction characteristics. Since it is hard to de-redden
6090 CBs one by one, this extinction correction was applied
statistically since we care about correcting trends in the PL
relation(s), not derivation of the real extinction affecting
individual CBs. The $A_J,A_H$, and $A_K$ extinction values were taken
from \citet{Rieke85}. These corrections have little influence on the
NIR magnitudes, but they are needed in the $V$ band. The best-fitting
period--color relations are
\begin{equation}\label{equation4}
  \begin{aligned}
   & (J-K_{\rm s})_0 = -1.19 \log P+(-0.16 \pm 0.02), \sigma_{JK_{\rm s}}=0.09,\\
   & (H-K_{\rm s})_0 = -0.20 \log P+(-0.02 \pm 0.01), \sigma_{HK_{\rm s}}=0.03,\\
   & (V-K_{\rm s})_0 = -4.14 \log P+(-0.36 \pm 0.08), \sigma_{VK_{\rm s}}=0.38,
   \end{aligned}
\end{equation}
and
\begin{equation}\label{equation5}
  \begin{aligned}
   & (J-K_{\rm s})_0 = 0.03 \log P+(0.17 \pm 0.02), \sigma_{JK_{\rm s}}=0.07,\\
   & (H-K_{\rm s})_0 = -0.00 \log P+(0.04 \pm 0.01), \sigma_{HK_{\rm s}}=0.04,\\
   & (V-K_{\rm s})_0 = -0.04 \log P+(0.72 \pm 0.07), \sigma_{VK_{\rm s}}=0.38,
   \end{aligned}
\end{equation}
for $\log( P \mbox{ day}^{-1}) < -0.25$ (Eq. \ref{equation4}) and
$\log( P \mbox{ day}^{-1}) > -0.25$ (Eq. \ref{equation5}),
respectively. These relations are almost independent of the PL
relations of Eq. (\ref{equation3}). A careful comparison shows that
both sets of equations above are in good mutual agreement, although a
correction to the early-type CB PL relation is required (as already
discussed in the context of the PL relations). We show the curves
  resulting from application of modern, non-parametric regression
  techniques in Figs 3--5, which were obtained using the standard
  Matlab {\tt ksrlin} and {\tt ksr} local linear kernel smoothing
  regression functions. For Figs 3 and 5 we used the default
  conditions, while for Fig. 4 we used a bandwidth or smoothing
  parameter $h=200$. This kind of approach is very useful when dealing
  with large-$N$ sample data without having to make `linear'
  assumptions; it allows us to detect non-linear features. This
  technique was applied to ensure that our linear fit is sufficient to
  cover the full distribution of our data points. The green dotted
  lines (in all three figures) are based on local linear regression
  techniques, while the black dash-dotted lines are based on smooth
  regression. These curves are comparable to our linear fits,
  considering the uncertainties, especially in the $H-K_{\rm s}$
  diagram. This implies that Eq. \ref{equation4} are reliable
  period--color relations.

\begin{figure}
\centering
\includegraphics[width=95mm]{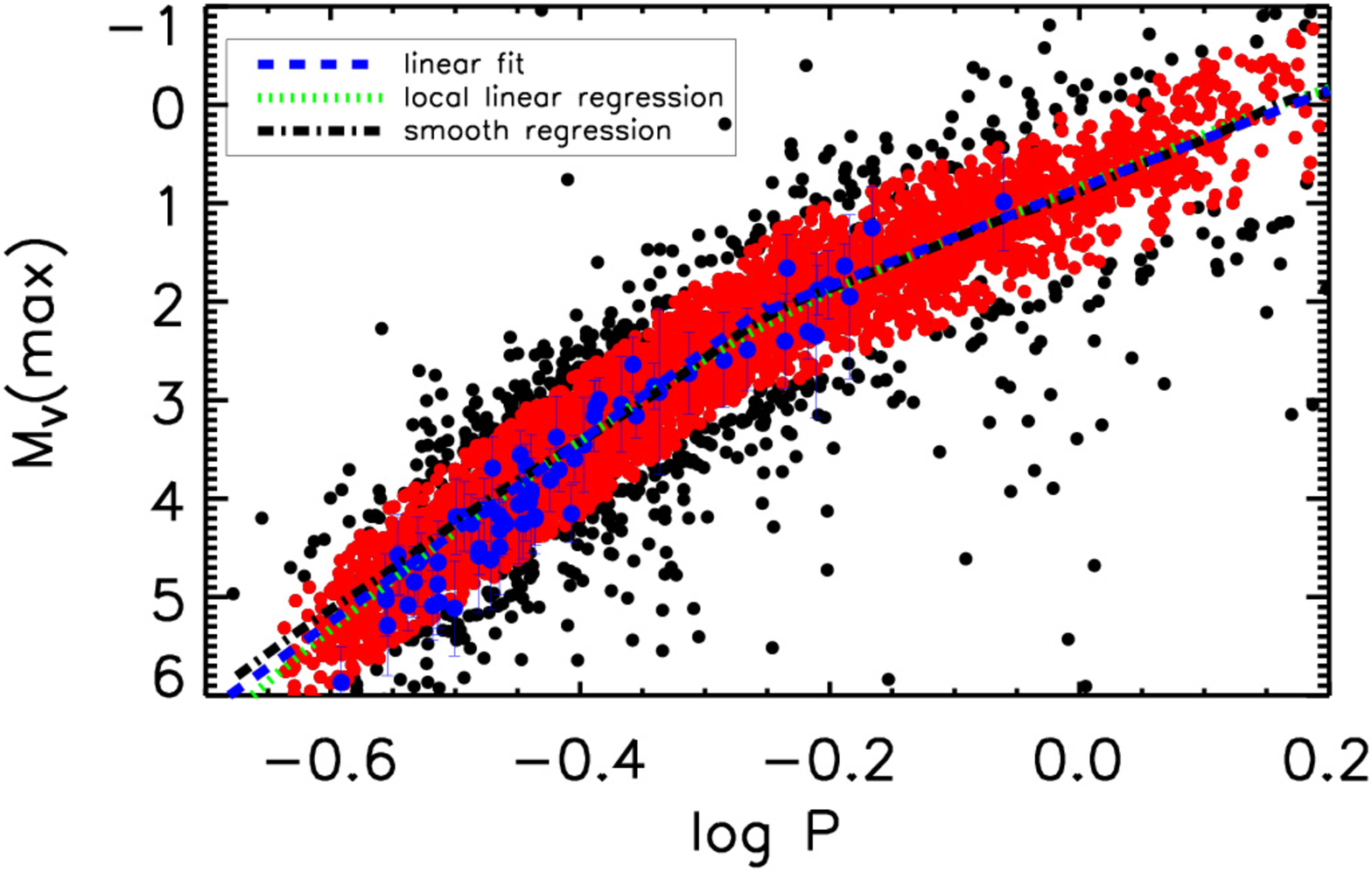}
\includegraphics[width=95mm]{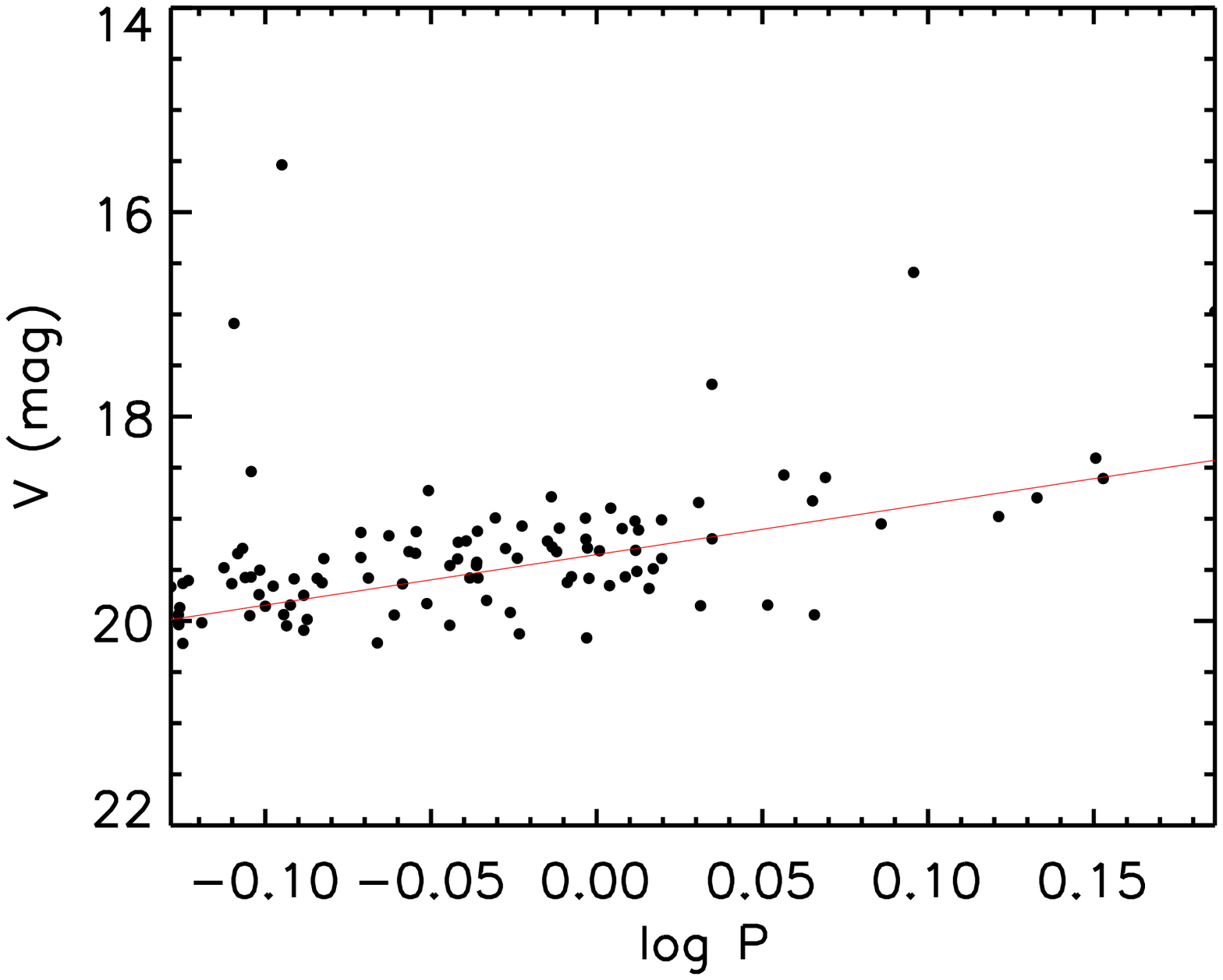}
\caption{$V$-band PL relations for (top) our full sample of 6090 CBs
  (black dots) and (bottom) the LMC CBs. Top: the PL relations for
  early- and late-type CBs are determined separately for periods
  shorter and longer than $\log( P \mbox{ day}^{-1}) = -0.25$. Blue
  dots: our sample of 66 calibration objects. Bottom: the red solid
  line is the $V$-band PL relation for Galactic CBs, Eq.
  (\ref{equation6}).\label{f9.fig}}
\end{figure}

From our newly established $JHK_{\rm s}$ PL relations, the $V$-band PL
relation can be derived using the NIR distances. We adopted $V_{\rm
  max}$ values from different literature sources, so large
uncertainties caused by calibration inhomogeneities are inevitable for
individual CBs. However, use of such a large sample makes their
combination statistically more reliable by averaging out calibration
differences, provided that the latter are not dominated by
unrecognized systematic errors. The resulting $V$-band PL relation
(see Fig. \ref{f9.fig}) is
\begin{equation}\label{equation6}
  \begin{aligned}
   & M_{V_{\rm max}} = (-9.15 \pm 0.12) \log P+(-0.23 \pm 0.05), \sigma_V=0.30 (\log P < -0.25), \\
   & M_{V_{\rm max}} = (-4.95 \pm 0.13) \log P+( 0.85 \pm 0.02), \sigma_V=0.35 (\log P > -0.25).
   \end{aligned}
\end{equation}

\section{Discussion}

\subsection{Assumptions and error budget}

  In Table \ref{t1} and Section 4, we explained how we obtained the
  maximum NIR magnitudes for all of our 66 calibration CBs. In this
  section, we discuss the underlying assumptions we had to make and
  the overall error budget. One important assumption is that the NIR
  reddening law is universal. Another major assumption is that the
  differential reddening across the face of each cluster is small, so
  that for a given OC CB we can adopt its host cluster's
  reddening. Since these OCs and CBs are affected by only little
  extinction---for most, $E(J-H)<0.1$ mag---the anticipated influence
  imposed by these effects can indeed be ignored. A third main
  assumption we made is that most CBs are only affected by small
  period changes, if any. We adopted the closest primary minimum
  epoch, $T_0$, to the observed 2MASS epoch to avoid any effects
  caused by period changes. Our final key assumption is that the
  light-curve shapes do not change significantly as a function of
  wavelength. These latter two assumptions will contribute 10\%
  uncertainties to both the total amplitude and the phase uncertainty
  for a given CB. The overall error includes the uncertainty
  associated with the distance modulus, $\sigma_{\rm DM}$, obtained
  using either parallaxes or the OC main-sequence fitting method, the
  photometric error, $\sigma_{\rm pho}$, the extinction-correction
  error, $\sigma_{\rm ext}$, and the uncertainty caused by converting
  single-epoch magnitudes to maximum magnitudes, $\sigma_{\rm
    conv}$. The maximum overall error $\sigma = \sigma_{\rm
    DM}+\sigma_{\rm pho}+ R_{\rm filter} \sigma_{\rm ext}+\sigma_{\rm
    conv}$, where $R_{\rm filter}$ is the total to selective
  extinction (i.e., the extinction law) for a given filter. These
  uncertainties are reported in Table \ref{t1}.

\subsection{Previous PLC or PL relations}

\citet{Rucinski94} derived a widely used PLC relation for late-type
CBs. Subsequently, \citet{Rucinski06} tried to derive a PL relation
based on 21 nearby, late-type CBs with accurate {\sl Hipparcos}
parallaxes; within the significant uncertainties, only 11 of his 21
objects followed the resulting PL relation, $V = (12.0 \pm 2.0) \log
P+(-1.5 \pm 0.8)$. Our PL relation is much shallower. This difference
can be easily understood. \citet{Rucinski94}'s PL relation was based
on only 10 objects, a sample where the CB with the shortest period (VW
Cep; $\log P=-0.55$) determined the slope. In our period--color
relation (Fig. \ref{f9.fig}), the slope is constrained by a large
number of intermediate-period CBs ($-0.5<\log P<-0.3$, in days) and
not by their short-period ($\log P<-0.55$) counterparts. In addition,
both short- and long-period ($\log P>0.1$) CBs are affected by
sampling incompleteness, which is caused by the obvious cut-offs at
$\log P=-0.6$ and $\log P=0.2$, respectively. The short-period cut-off
is driven by the long formation timescale of faint CBs (longer than
the age of the Milky Way), while long-period CBs easily lose their
angular momentum because of the convective nature of the stellar
surfaces. The latter determines the upper limit to the initial stellar
mass and, hence, the upper limit to the period. To correct the
\citet{Rucinski06} PL relation for potential selection effects, 
  we applied weights to his data as a function of orbital period,
  using the magnitude uncertainty. We plotted the orbital-period
  distribution in 0.01 dex bins, and we calculated the probability
  $\xi_i$ in each bin. If $n$ of \citet{Rucinski06}'s CBs are located
  in a given bin, the probability weights of these CBs are
  $\frac{\xi_i}{n}$. We next consider the uncertainties, $\sigma_M$,
  in the absolute magnitudes of these CBs. The final weight is
  $\frac{\xi_i}{n {\sigma_M}^2}$, for each CB. The revised PL
relation based on the \citet{Rucinski06}'s data thus corrected is $V =
(-11.4 \pm 1.7) \log P+(-1.2 \pm 0.8)$ mag, which is indeed close to
our result. If we exclude TY Men, the newly revised PL relation
becomes $V = (-10.5 \pm 2.2) \log P+(-0.9 \pm 1.0)$ mag, which is even
closer to our result.

\subsection{NIR PL relation accuracy}

Most recent studies using Cepheids as distance tracers are based on
NIR PL relations. They are more accurate and significantly less
sensitive to extinction and metallicity variations than $V$-band PL
relations. In this paper, we derived the first PL relations for
bright, early-type CBs, which allows the application of CBs to trace
features at greater distances compared with the use of the late-type
CB PL relation. This study has thus made CBs a viable distance tracer
pertaining to old environments. Combining the $JHK_{\rm s}$ PL
relations, we derived the distances to our full sample of 6090
CBs. The resulting accuracy is high: 90\% of our sample CBs have
distance errors of less than 5\%, and 95\% have distance uncertainties
of less than 10\%. The remaining 5\% may be CBs with poor-quality
photometry, CBs affected by complicated differential extinction (or
severe reddening), or objects that have been misidentified as
CBs. The latter include small-amplitude RR Lyrae and ellipsoidal
  binaries.

  We will now evaluate the accuracy in distance as determined by
  our NIR PL relations. By combining all three NIR PL relations, the
  statistical uncertainty will decrease to 0.05 mag. The systematic
  error includes the zero-point errors in the PL relations, the
  reddening error, the metallicity error, and the photometric
  uncertainties. Since our PL relation relies on the distances to our
  sample OCs and on the parallaxes, zero-point errors were introduced
  by the method we used for distance determination. The
  distance-modulus uncertainties of the 17 sample OCs for which we
  obtained distances in this manner (see Table 1) are 0.1--0.2 mag and
  the average distance-modulus error associated with the 20 CBs in our
  sample for which we have access to {\sl Hipparcos} parallaxes is
  0.137 mag. The average of the combined sample of 18 distances is
  $\sigma=0.138$ mag and so the error in the zero-point is
  $\frac{\sigma }{\sqrt{n}}=0.032$ mag. As for the reddening
  uncertainty, the NIR reddening error is very small compared with
  those in optical bands, with $A_J/A_V=0.282$, $A_H/A_V=0.175$,
  $A_K/A_V=0.112$ \citep{Rieke85}. More than half of our CBs are
  nearby CBs, and they have a reddening corresponding to $E(B-V)<0.08$
  mag. Even if the $R_V=3.1$ reddening law would need to be adjusted
  by 25\%, the NIR reddening error would still be less than 0.01
  mag. Although the reddening may still contribute to the statistical
  error, it is not a significant contributor to the systematic error.

\begin{figure}
\centering \includegraphics[width=95mm]{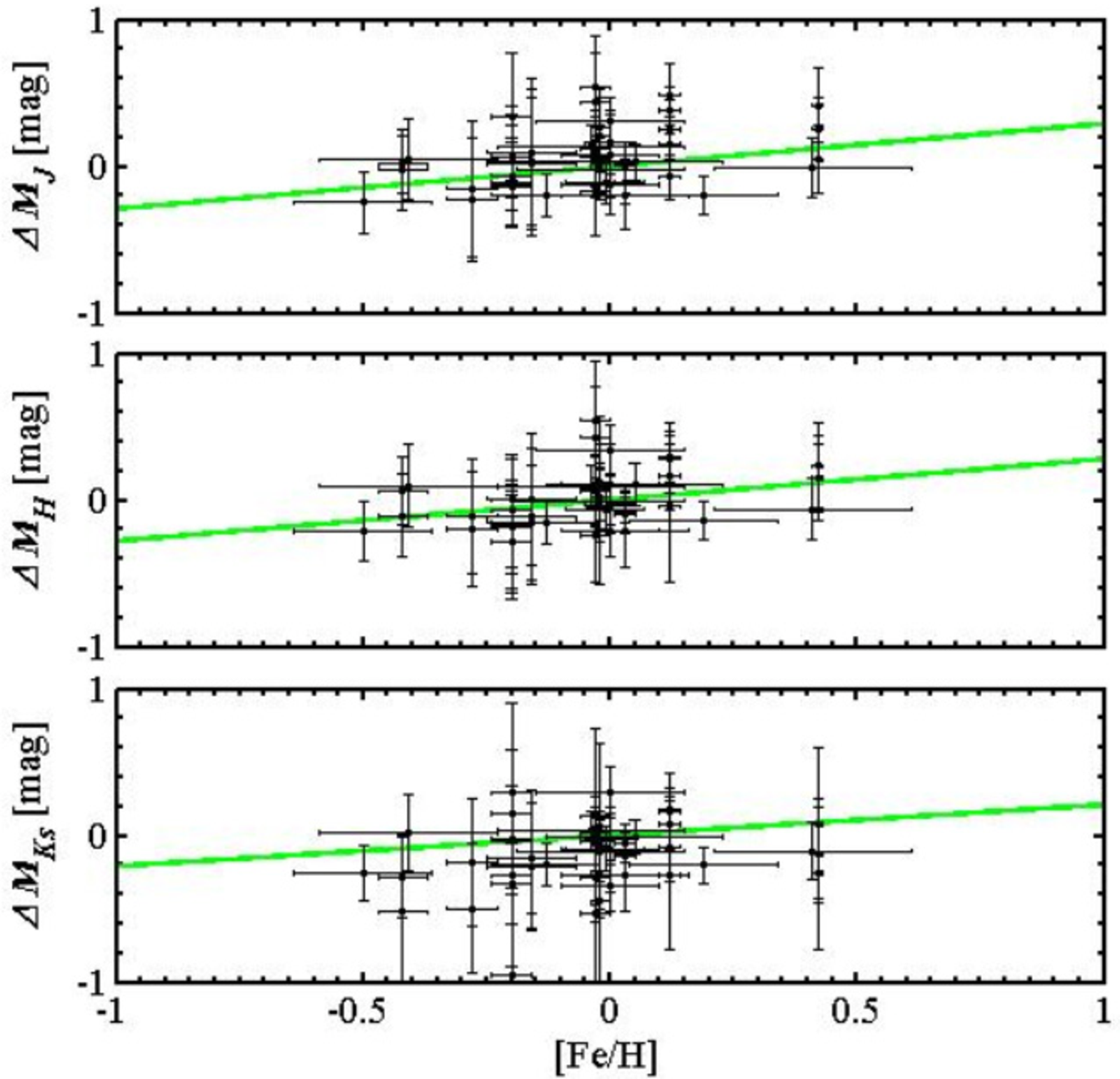}
\caption{NIR PL relation magnitude residuals as a function of
  metallicity. The green solid lines are linear fits.\label{f11.fig}}
\end{figure}

  The metallicity error is caused by the different metallicities of
  the CBs used to derive the PL relation, which could potentially
  result in a PL relation that depends on metallicity. To evaluate
  this possibility, a relation between $\Delta M$ and [Fe/H] must be
  established, where $\Delta M$ is the deviation of the observed
  absolute magnitudes to the predicted absolute by PL
  relation. Fortunately, for 48 of our 66 calibration CBs we have
  access to direct metallicity information \citep{Rucinski13} or to
  metallicity measurements obtained for their host OCs
  \citep{Dias02}. All of the latter metallicities were obtained from
  the latest available literature sources. The NIR $\Delta M$--[Fe/H]
  relations are $M_J=0.29\rm{[Fe/H]}\pm0.14$ mag,
  $M_H=0.28\rm{[Fe/H]}\pm0.12$ mag, and $M_{K_{\rm
      s}}=0.21\rm{[Fe/H]}\pm0.15$ mag (see Fig. \ref{f11.fig}). This
  means that if CBs with $[Fe/H]=-1.0$ dex are used to derive the PL
  relations, a systematic uncertainty of at least 0.2 mag may be
  applicable. The average metallicity of the 48 CBs is
  $\overline{\rm{[Fe/H]}}= -0.041$ dex, so the systematic error
  introduced by metallicity differences is around $\sigma=0.01$
  mag. The total systematic error is $\sigma_{\rm sys}^2=\sigma_{\rm
    zp}^2+\sigma_{\rm ext}^2+\sigma_{\rm metal}^2$, i.e., $\sigma_{\rm
    sys}=0.03$ mag. As a result, the accuracy of using CBs as distance
  tracers is 0.05 (statistical)$\pm$0.03 (systematic) mag.

\subsection{distance to LMC}

\citet{Graczyk11} published a catalog of 26,121 eclipsing binary stars
in the LMC, identified based on visual inspection of the Optical
Gravitational Lensing Experiment III catalog. Their 1048 type-EC
eclipsing binaries are CBs. In fact, based on their light curves, most
are semi-detached binaries with unequal minima and other types of
variable stars. They only included CBs with long periods ($\log P
>-0.2$). To select CBs that can be used as distance tracers, we
adopted the period--color selection of Eq. (\ref{equation6}) and
Fig. \ref{f5.fig}. We derived a color cut at $(V-I)_0 = (0.41 \pm
0.21)$ mag using the transformation equations of \citet{Bessell98}. We
imposed a period selection of $-0.13<\log P <0.2$, where the upper
limit is at the long-period end of the CB distribution and the lower
limit is the magnitude limit for detecting LMC CBs. This led to a
final sample of 102 LMC CBs, resulting in a distance modulus of
$(m-M_{V})_0=18.41\pm 0.20$ mag. This is first distance to the LMC
that was determined based on CBs. It is fully consistent with the
current best LMC distance modulus \citep{de Grijs14}, $(m-M)_0=18.49
\pm 0.09$ mag. Figure \ref{f9.fig} shows that the LMC CBs follow the
same PL relation as their counterparts in our Galaxy. Previous studies
of large samples of eclipsing binary systems in the LMC
\citep{Muraveva14, Pawlak16} did not find a clear PL relation, since
they did not correct for the prevailing period--color relations.

\section{Summary}

We collected CBs in OCs and CBs with accurate {\sl Hipparcos}
parallaxes. Our full sample contains 6090 CBs from the GCVS and ASAS
surveys, while the OC sample contains 2167 OCs. To exclude foreground
and background CBs, (i) the CB of interest must be located inside the
core radius of its host OC; (ii) the CB's proper motion must be
located within the $2\sigma$ distribution of that of its host OC; and
(iii) the CB's age must be comparable to that of its host OC, $\Delta
\log (t\mbox{ yr}^{-1}) <0.3$. We thus selected 42 high-probability OC
CBs. Combined with four nearby moving-group CBs and 20 W UMa-type CBs
with accurate {\sl Hipparcos} parallaxes, a sample of 66 CBs is used
to determine the $JHK_{\rm s}$ PL relations. The latter yield
distances that are as accurate as those resulting from the $JHK_{\rm
  s}$ Cepheid PL relations ($\sigma < 0.10$ mag).

NIR PL relations for early-type CBs are obtained for the first
time. To check the reliability of our PL relations, the CB
period--color relations are also investigated. These can be used to
exclude unreliable CBs. We discuss the potential of CBs as distance
tracers and find that they can determine distances to 5\% uncertainty
for 90\% of the objects in our full sample. We also discuss the
  overall uncertainty associated with using CBs as distance tracers
  and we derive a value of $\sigma=0.05 \mbox{ (statistical)} \pm0.03
  \mbox{ (systematic)}$ mag. The 102 CBs in the LMC satisfying our
period--color selection are used to determine an LMC distance modulus
of $(m-M_V)_0=18.41\pm0.20$ mag.

  Since more than 30,000 CBs have been found in the GCVS, ASAS,
  OGLE III, and Catalina catalogs, and hundreds of additional CBs are
  reported on by individual studies every year, CBs can be important
  tracers to study the structure of our Galaxy. However, given the NIR
  relations determined here, more accurate NIR absolute magnitudes
  should be obtained for a large sample of CBs to make this aim a
  reality.  Combined with information about their periods, physical
  parameters from light-curve solutions, and age information from the
  host clusters, the formation and evolution mechanisms of CBs could
  indeed be much better understood. {\sl Gaia} will observe most known
  CBs; distances derived based on our PL relations can hence be used
  as a cross check of the {\sl Gaia} parallaxes anticipated soon.

\acknowledgments{We thank acknowledge useful comments from an
  anonymous referee. We are grateful for research support from the
  National Natural Science Foundation of China through grants 11373010
  and 11473037.}


\end{document}